\documentstyle[twoside,fleqn,npb,epsfig]{article}
\newcommand{\AmS}{{\protect\the\textfont2
  A\kern-.1667em\lower.5ex\hbox{M}\kern-.125emS}}

\title{Progress in calculating hexagon amplitudes at one--loop}

\author{T.~Binoth\address{University of Edinburgh, School of Physics,\\
            EH9 3JZ Edinburgh, Scotland UK}}%

\begin{document}

\begin{abstract}
In this article I review recent progress towards the 
analytical calculation of massless 6--point amplitudes.
A way to organize such calculations is sketched and
results for  scattering amplitudes in the Yukawa model are 
presented.
\end{abstract}

\maketitle

\section{Introduction}
The high energy experiments of the next ten years are 
hadron colliders operating at the multi--TeV scale. 
Because of the high center--of--mass energies at 
the Tevatron and the LHC,
the final states will look very complex.
One may say that we are entering the decade of 
{\em multi}--particle/jet physics.

The expected jet rates at the LHC are huge.
A leading order analysis
is not adequate to make a detailed prediction for these
cross sections.  To address the problems with a leading order description
three points  can be made:

1) Large scale dependence: An N--jet cross section behaves like 
$\alpha_s(\mu)^N$, means 
scale uncertainties are amplified with the number of jets.

2) Peripheral phase space regions:
Partonic cross sections are very sen\-si\-ti\-ve to next--to--leading
order effects, when severe cuts are applied, 
which is typical for background
processes.

3) Jet structure:
The more information about the matrix elements is known
the better the internal structure of a jet can be described.

It is important to note that the ability
to detect New Physics at the LHC and Tevatron crucially 
depends on the understanding of the corresponding Standard Model
backgrounds. All that is
sufficient motivation to treat the generic multi--particle/jet
final states at the next--to--leading order (NLO) level.

Presently our computational skills for describing 
N--jet production at NLO end already at N=3.
The pioneering work to calculate the amplitudes
for two--jet processes at NLO in hadronic collisions
was accomplished by Ellis and Sexton in 1986 \cite{Ellis:1985er}.
The partonic amplitudes relevant 
for 3 jet production were provided by Bern, Dixon and Kosower 
\cite{Bern:1993mq}
and Kunszt, Signer, Tr\a'ocs\a'anyi \cite{Kunszt:1994tq} in 1993/94.
New technology was invented for these calculations
borrowed partly from string theory. Further, supersymmetry relations
were exploited and helicity methods were applied
at the NLO level \cite{Dixon:1996wi}. Other needed ingredients were
reduction formulas for Feynman parameter integrals 
with nontrivial numerators and representations of scalar
5--point functions (pentagon integrals) \cite{Bern:1993kr}. 
For QCD the step to NLO 6--point amplitudes has not been
taken yet, albeit its phenomenological relevance
for collider physics. 

\section{Scalar scattering in the Yukawa model}
To understand the organization of hexagon amplitudes and 
their efficient computation it is necessary to investigate 
examples which do not exhibit the full complexity 
of a generic gauge theory amplitude. An amplitude 
which contains only scalars as external particles 
is likely to be a good candidate.
Let us consider 6--scalar scattering 
in the Yukawa model.
\begin{eqnarray}
{\cal L}_{\rm yuk}
=
\bar\psi
\Bigl[
i\,\partial\!\!\!/
-m 
-g\phi
\Bigr]
\psi
+ {\cal L}_{\rm \phi}
\label{defyukawamodel}
\end{eqnarray}
We restrict ourselves to the massless case
and the fermion loop contribution. The corresponding amplitude 
is given by
\begin{eqnarray}\label{Eq:start}
\Gamma^{\phi}_{\rm yuk}[p_1,p_2,p_3,p_4,p_5,p_6] = \hspace{2.5cm} \nonumber\\
 \frac{-g^6}{(4\pi)^{n/2}6}
 \sum\limits_{\pi\in S_6} {\cal A}(p_{\pi_1},p_{\pi_2},
                  p_{\pi_3},p_{\pi_4},p_{\pi_5},p_{\pi_6})
\end{eqnarray} 
Each permutation corresponds to a single Feynman diagram. The amplitude 
for the trivial permutation is given by 
\begin{eqnarray}\label{amp_graph}
{\cal A}(p_1,p_2,p_3,p_4,p_5,p_6) = \hspace{1.5cm} \nonumber\\
\int \frac{d^nk}{i \pi^{n/2}}
\frac{{\rm tr}(\not q_1\not q_2\not q_3\not q_4\not q_5\not q_6)}{q_1^2q_2^2q_3^2q_4^2q_5^2q_6^2}
\end{eqnarray}
where $q_j = k - r_j = k-p_1-\dots-p_j $.
Working out the spinor trace one ends up with
scalar products of loop and external momenta. 
One can show that all scalar products which contain
loop momenta can be expressed  as inverse propagators
which  can be canceled directly. In this way one is getting
rid of all loop momenta in the numerator, or in other words:
The Yukawa amplitude for multi--scalar scattering 
has a trivial tensor structure.
A given graph of the scalar hexagon amplitude decays immediately
into scalar 3,4,5,6--point integrals. 

The representation of the amplitude in terms of these
scalar integrals is not useful as spurious infrared poles
are present which have to cancel
in the full amplitude. One can show that none of the graphs is
actually infrared divergent. The tool to make the infrared cancellations
manifest are scalar reduction formulas.
The N--point n--dimensional scalar integral integral in 
Feynman parameter space
is given by
\begin{eqnarray}\label{scalar_integral}
I_N^n(p_1,\dots,p_N) = (-1)^N \Gamma(N-n/2) \hspace{1cm} \\ 
 \int_{0}^{\infty} d^Nx 
\frac{\delta(1-\sum_{l=1}^N x_l)}{(x\cdot S \cdot x/2)^{N-n/2}} 
\nonumber
\end{eqnarray}
and the following reduction formula holds
\begin{eqnarray}\label{EQscalarreductionNT}
I_N^n &=& \sum \limits_{k=1}^{N} B_k \,
  I_{N-1,k}^n  
  \nonumber\\ 
  &&- (N-n-1) \,\sum \limits_{k=1}^{N} B_k\, I_N^{n+2}  
\end{eqnarray}
$S$ is the matrix which carries the kinematic information.
Its entries are Mandelstam variables $s_{ij\dots}=(p_i+p_j+\dots)^2$
formed by the external momenta $p_1,\dots,p_N$.
$I_{N-1,k}^n$ stands for the $k$th pinch integral of $I_N^n$,
means one leaves out the $k$th propagator in the momentum representation.
The reduction coefficients $B_k$ are the solutions of 
a simple linear equation $\sum_{l=1}^{N} S_{kl} B_l = -1$.
They will play a special role in what follows. They can
be written  very compactly in terms
of spinor traces. For example
\begin{eqnarray}
B_1=
\frac{ {\rm tr}(123456){\rm tr}(3456)
             -2 s_{34}s_{45}s_{56} {\rm tr}(6123) }{
	4 s_{12}s_{23}s_{34}s_{45}s_{56}s_{61} - {\rm tr}(123456)^2}\nonumber
\end{eqnarray}  
and the other B's are obtained by cyclic permutations.
Here ${\rm tr}(12\dots)$ is a shorthand for 
${\rm tr}(\not p_1\not p_2\dots)$, for details see \cite{Binoth:2001vm}.

Applying the reduction formulas allows to express
each Feynman diagram in terms of
triangle graphs with 0,1 or 2 light--like
external momenta and 6--dimensional box functions.
Only the triangle graphs with light--like legs carry
infrared poles. 
As each individual Feynman diagram is infrared 
finite, these triangle graphs  have to cancel.
The cancellations of the infrared divergent triangles 
gave rise to very interesting insights. 
There are a number of nonlinear constraints between Mandelstam variables
which are represented {\em linearly}
in terms of the reduction coefficients, e.g. one finds
\begin{eqnarray}
{\rm tr}(3456) + {\rm tr}(123456)  B_1 + 2 s_{34}s_{45}s_{56}  B_4 = 0
\end{eqnarray}
This made the whole structure of the calculation very 
transparent and an astonishingly compact result for the Yukawa
6--scalar amplitude could be derived \cite{Binoth:2001vm}.  It was found
\begin{eqnarray}\label{Eq:final}
\Gamma_{\rm yuk}^{\phi}[p_1,p_2,p_3,p_4,p_5,p_6] = \hspace{2.cm} \nonumber \\
 - \frac{g^6}{(4\pi)^2}\sum\limits_{\pi\in S_6}^{} G(p_{\pi_1},p_{\pi_2},
                  p_{\pi_3},p_{\pi_4},p_{\pi_5},p_{\pi_6})
\end{eqnarray} 
with\begin{eqnarray} G(p_1,p_2,p_3,p_4,p_5,p_6) = 
 \frac{2}{3} \,I^n_3(p_{12},p_{34},p_{56}) \nonumber \\
+\Bigl\{
 B_1\frac{[ {\rm tr}(6123)-2s_{61}(s_{123}-s_{12}) ]}{s_{123}s_{345}-s_{12}s_{45}} 
 \nonumber \\
+B_2\frac{[ {\rm tr}(1234)-2s_{34}(s_{123}-s_{23}) ]}{s_{234}s_{123}-s_{23}s_{56}} 
 \Bigr\} 
\nonumber \\ \times
\log\left( \frac{s_{12}}{s_{123}} \right)
         \log\left( \frac{s_{23}}{s_{123}} \right) \nonumber \\
-\Bigl\{ B_1 - 
B_2\frac{\left[ {\rm tr}(1234) - 2 s_{34} (s_{123}-s_{23}) \right]}{2
[s_{234}s_{123}-s_{23}s_{56}]} 
\nonumber \\
- B_6\frac{\left[ {\rm tr}(5612)- 2 s_{56} (s_{345}-s_{61})\right]}{2 
[s_{345}s_{234}-s_{61}s_{34}]}
\Bigr\}\times
\nonumber \\ 
\Bigl[ 
       \log\left( \frac{s_{12}}{s_{234}} \right) 
       \log\left( \frac{s_{56}}{s_{234}} \right)\nonumber\\ 
     + \log\left( \frac{s_{34}}{s_{234}} \right) 
       \log\left( \frac{s_{12}}{s_{56}} \right) \Bigr]\nonumber   
       \label{ampsi}
\end{eqnarray}
The result consists only of triangle functions
and $\log\log$ terms. Apart from dilogarithms
hidden in the finite triangle function no  
dilogarithms are present in the result. This can be understood
by the fact that the dilogarithms which are present at
intermediate steps can be related to single pole
contributions from triangle graphs with one light--like
leg and is infrared divergent. 
As the infrared divergence vanishes, the dilogarithms also have to vanish.
The remarkably simple form of the amplitude
could be achieved because it was understood that the 
reduction coefficients $B_j$ are the natural building 
blocks of the amplitude.  This lead to the {\em conjecture} that 
massless hexagon amplitudes in general gauge theories 
can be represented most compactly as some linear combination of 
reduction coefficients. 
Still the conjecture has to be if not proven
then at least confirmed by other calculations which exhibit
a non--trivial tensor structure. 

\section{$\gamma\gamma\to 4$ scalars in the Yukawa model}
To study an amplitude with a non--trivial tensor structure
we considered the gauged version of Eq.~(\ref{defyukawamodel}).
For example the fermion loop contribution to 
the 
\begin{equation}
\gamma(p_1)\gamma(p_2)\to \phi(-p_3)\phi(-p_4)\phi(-p_5)\phi(-p_6)
\end{equation} 
amplitude  is
given by a sum over permutations of Feynman diagrams 
which differ only by the position of the photons and scalars
at the fermion loop.
\begin{eqnarray}
\Gamma^{\lambda_1\lambda_2} 
= -\frac{g^4e^2}{(4\pi)^{n/2}} {\cal M}^{\lambda_1\lambda_2}
=-\frac{g^4e^2}{(4\pi)^{n/2}} \nonumber\\
\times\frac{1}{6} \sum\limits_{\pi\in S_6} {\cal A}^{\lambda_1\lambda_2}(p_{\pi_1},p_{\pi_2},
                  p_{\pi_3},p_{\pi_4},p_{\pi_5},p_{\pi_6})
\end{eqnarray}
with, e.g.
\begin{eqnarray}\label{Eq:ampbygraphs}  
{\cal A}^{\lambda_1\lambda_2}(p_{1},p_{\pi_2},
                  p_{\pi_3},p_{\pi_4},p_{\pi_5},p_{\pi_6})=\nonumber\\
\int \frac{d^nk}{i\pi^{n/2}}
\frac{\mbox{tr}(\not\varepsilon_1^{\lambda_1}\not q_1
                \not\varepsilon_2^{\lambda_2} \not q_2\not q_3\not q_4\not q_5\not q_6)}
		{q_1^2q_2^2q_3^2q_4^2q_5^2q_6^2} \nonumber	  
\end{eqnarray}
Here $\lambda_j$ are the helicities of the photons.
$q_j=k-r_j=k-p_1-\dots -p_j$ are the propagator momenta.
The amplitude exhibits an $S_4$ Bose symmetry, as
the scalars are freely interchangeable. 
For a given ordering of the scalars one finds 20
different Feynman diagrams corresponding to the different possibilities
to attach the two photons. 
This set forms a gauge
invariant sub--set of all Feynman diagrams.
It is sufficient to consider this sub--set and performing
the summation over remaining permutations explicitly. 
\subsection*{Gauge and Lorentz structure}
To cope with the combinatorial complexity of hexagon amplitudes
it is useful to split the problem into irreducible pieces
as far as possible. The guidelines are the analyticity and
Lorentz structure of the amplitude together with gauge invariance.
The external momenta $p_1,\dots,p_6$ obey momentum conservation. 
To make this explicit
we introduce the five vectors $r_j=p_1+\dots+p_j$, $j = 1,\dots,5$.
We assume for the moment that the kinematics is non--exceptional, i.e.
the 5 vectors define a 5--dimensional
subspace of the n--dimensional space. 
The scattering tensor ${\cal M}$ has the following general decomposition
\begin{eqnarray}\label{ABs}
{\cal M}^{\mu_1 \mu_2} &=& 
        A \, g^{\mu_1\mu_2} 
+\sum\limits_{j_1,j_2=1}^{5} 
  B_{j_1 j_2}\,r_{j_1}^{\mu_1}\,r_{j_2}^{\mu_2}
\end{eqnarray}
This representation contains 26 coefficients.
The transversality conditions 
$\varepsilon_1\cdot p_1 = 0$, 
$\varepsilon_2\cdot p_2 = 0$  
reduce the number of relevant coefficients to 17.
Further, by gauge invariance
\begin{equation}\label{ward_id}
{\cal M}^{p_1\varepsilon_2} = {\cal M}^{\varepsilon_1 p_2} = 0
\end{equation}
which leads to 7 independent linear relations between the
tensor coefficients. They may be used to express 
all 17 coefficients in terms of 
10 unconstrained coefficients, e.g. $A$, $B_{ij}$, $i,j=3,4,5$.
It follows that one has to calculate only these 10 independent
coefficients. Note that gauge invariance is completely exploited
in this way. On the other hand, calculating more coefficients and confirming
the Ward identities explicitly provides a  test of the 
calculation. 
By using helicity methods directly one also sees that
10 tensor coefficients are sufficient to determine the
full amplitude. 

The problem is now reduced to the computation of 10 
tensor coefficients for 20 6--point Feynman diagrams.

\subsection*{Projection techniques}
The tensor coefficients in Eq.~(\ref{ABs})
can be determined by acting with projection operators onto
each Feynman diagram. 
This may be achieved by defining 
\begin{eqnarray} 
{\cal P}^{\mu\nu} &=& \frac{1}{n-5} \left(g^{\mu\nu} 
  - \sum\limits_{j,k=1}^{5} 2 r_j^{\mu}  H_{jk} r_k^{\nu} \right) \\
{\cal R}_j^{\mu}  &=& 2 \sum\limits_{k=1}^{5}  H_{jk} r_k^{\nu}  
\end{eqnarray}
$H$ is the inverse of the $5\times 5$ Gram matrix defined by 
$G_{jk} = 2 r_j\cdot r_k$.
Note that the inverse of $G$ is only defined if $n\neq 4$.
This will not lead to a problem as one can
cancel the inverse matrices before the limit $n\rightarrow 4$
has to be taken.

One may easily check the following relations for the tensors 
${\cal P}^{\mu\nu}$ and ${\cal R}_j^\mu$:
\begin{eqnarray}
{\cal P}^{\mu\rho} {\cal P}_{\rho}^{\;\;\nu} &=& {\cal P}^{\mu\nu}  \\
{\cal P}^{\mu}_{\;\;\nu} p_j^{\nu} &=& 0\\
{\cal P}^{\mu}_{\;\;\mu} &=& tr({\cal P}) = tr({\cal P}\cdot {\cal P}) 
= 1 \\
{\cal P}^{\mu}_{\;\;\nu} {\cal R}_j^{\nu} &=& 0\\
{\cal R}_{j,\nu} r_k^{\nu} &=& \delta_{jk} \\
{\cal R}_{j,\nu} {\cal R}_{l}^{\;\nu} &=& 2\, H_{jl}
\end{eqnarray}
${\cal P}$ is a projector onto the $(n-5)$ dimensional subspace
perpendicular to the 5 dimensional space spanned by the 
vectors $r_j$. ${\cal R}_{j,\nu}$ 
is the dual vector to $r_j^{\nu}$ relative to this $(n-5)$--dimensional space.
Here and in the following we use summation conventions for Lorentz 
and vector indices.

To determine the tensor coefficients, $A$ and 
$B_{kl}$, we define the following linear operators 
\begin{eqnarray}
\tilde A({\cal M}) &=& {\cal P}_{\mu_1\mu_2} 
{\cal M}^{\mu_1 \mu_2}\nonumber\\
\tilde B_{kl}({\cal M}) &=& \Bigl[
{\cal R}_{k\mu_1}{\cal R}_{l\mu_2} - 2 \, H_{kl}\,{\cal P}_{\mu_1\mu_2}
\Bigr] {\cal M}^{\mu_1 \mu_2} \nonumber
\end{eqnarray}
The action of dual vectors ${\cal R}_l^\mu$ 
and projectors ${\cal P}^{\mu\nu}$ on Lorentz indices 
induces a natural mapping
from the momentum to the Feynman parameter representation
of the Feynman diagrams. 
When acting on Feynman diagrams one finds for example that
\begin{eqnarray}
\int \frac{d^nk}{i\pi^{n/2}}
    \frac{k \cdot {\cal P}\cdot k}{q_1^2\dots q_N^2} 
&=& -\frac{1}{2} \, I_N^{n+2} \\
\int \frac{d^nk}{i\pi^{n/2}}
    \frac{k \cdot {\cal R}_j }{q_1^2\dots q_N^2} 
&=& I_N^{n}(j) 
\end{eqnarray}
where the parameter integrals with non--trivial numerators 
are defined by \cite{Binoth:1999sp}
\begin{eqnarray}
I_N^n(j_1,\dots,j_R) = 
(-1)^N \Gamma(N-n/2) 
\nonumber\\
\int_{0}^{\infty} d^Nx 
\frac{\delta(1-\sum_{l=1}^N x_l)x_{j_1}\dots x_{j_R}}{(x\cdot S \cdot x/2)^{N-n/2}} 
\end{eqnarray} 
The reduction formulas necessary to express these integrals in terms
of scalar integrals in shifted dimensions can be found in 
\cite{Bern:1993kr,Binoth:1999sp,Heinrich:2000kj}.
After applying the reduction formula
the tensor coefficient of each Feynman graph is expressed 
in terms of scalar integrals with trivial numerators in shifted
dimensions similar to the approach of \cite{Fleischer:1999hq}.

\subsection*{Bases of scalar integrals}
After having reduced the scalar integrals with non--trivial
numerators each graph is written as a linear combination of
scalar 2-- to 6--point integrals 
in dimensions that are ranging from $n$ to $n+4$.
By applying the 
scalar reduction formula Eq.~(\ref{EQscalarreductionNT})
and explicit formulas for $(n+2m)$--dimensional triangle
functions with light--like legs 
these integrals can be mapped to
\begin{eqnarray}
I_2^n, I^{n}_{3, 1\&2 mass}, I^{n}_{3, 3 mass},
I_4^{n+2},I_5^{n+2} 
\end{eqnarray} 
Now each Feynman diagram is decomposed into a linear combination
of these functions. As each of the graphs
is IR/UV finite, all IR/UV poles have to vanish.
It turns out that the coefficients of
$I_2^n$, $I^{n}_{3, 1\&2 mass}$, $I^{n+2}_{3, 1 mass}$ cancel as a whole
in the amplitude.
As we have proven in \cite{Binoth:1999sp} that
all higher dimensional integrals with more than four 
external legs have to vanish also the $I_5^{n+2}$ part has to vanish
on a graph by graph bases. Note that the sorting into different scalar integrals
is also a sorting into different analyticity classes, as different
scalar integrals have different cut structures. No
cancellations are to be expected between these classes
and indeed are not observed. 

\subsection*{Result}
With the procedure outlined above all tensor coefficients 
were calculated. The results for these are lengthy
and not presentable.   
When combining them to  form the actual 
$++$ and $+-$ helicity amplitudes many interesting
cancellations were observed which are due to the same relations
that led to a compact result in the 6--scalar case. 

We find for the $++$ amplitude 
\begin{eqnarray}
{\cal M}^{++} =\frac{\varepsilon_1^+\cdot \varepsilon_2^+}{s_{12}}
\sum\limits_{\pi\in S_4} 
\Bigl\{
C(1,\pi_3,2,\pi_4,\pi_5,\pi_6)\nonumber\\ 
\times\Bigl[ 
      F_1( s_{\pi_4\pi_5},s_{\pi_5\pi_6},s_{12\pi_3} )\nonumber\\
+     F_1( s_{1\pi_3},s_{2\pi_3},s_{12\pi_3} )\nonumber\\
-2 \, F_1( s_{1\pi_3},s_{3\pi_4},s_{1\pi_3\pi_4} )\nonumber\\
+2 \,F_{2}( s_{12\pi_3}, s_{2\pi_4\pi_5},s_{1\pi_3},s_{\pi_4\pi_5}) \nonumber\\
-  F_{2}( s_{1\pi_3\pi_4}, s_{2\pi_4\pi_5},s_{1\pi_3},s_{2\pi_5}) \nonumber\\
-  F_{2}( s_{1\pi_3\pi_4}, s_{2\pi_3\pi_4},s_{\pi_3\pi_4},s_{\pi_5\pi_6})
\Bigr] \nonumber\\
+2 \, F_{2}( s_{1\pi_3\pi_4}, s_{2\pi_3\pi_4},s_{\pi_3\pi_4},s_{\pi_5\pi_6})
\nonumber\\
 + ( 1 \leftrightarrow 2 ) \Bigr\} 
\end{eqnarray} 
with
\begin{eqnarray}
C(1,2,3,4,5,6) = 2 \hspace{2.8cm}\nonumber\\
+\frac{\mbox{tr}^+(3456) B_5 + \mbox{tr}^+(1234) B_1}{s_{34}}\nonumber\\
+\frac{\mbox{tr}^+(6123) B_2 + \mbox{tr}^+(4561) B_4}{s_{16}}
\end{eqnarray} 
and 
\begin{eqnarray}
F_1(s_1,s_2,s_3) = -\log\left(\frac{s_1}{s_3}\right)\log\left(\frac{s_2}{s_3}\right) \nonumber\\
F_{2}(s_1,s_2,s_3,s_4) =  -\frac{1}{2}\log^2\left(\frac{s_3}{s_4}\right) \hspace{0.5cm}\nonumber\\
   -\mbox{Li}_2\left(1-\frac{s_1s_2}{s_3s_4}\right)
\end{eqnarray}
Here $\mbox{tr}^+(j\dots)=\mbox{tr}((1+\gamma_5)\, p_j \!\!\!\!\!/\; \dots)/2$.
The amplitude has indeed a simple representation in terms
of reduction coefficients as conjectured above.
\section{Conclusion}
It was pointed out that NLO corrections to multi--particle/jet
production at future colliders should be included to have reliable
theoretical predictions. On the way to understand
the organization of 6--point processes at the one--loop level
analytical results have been presented for 6--scalar scattering in the
Yukawa model. Very compact expressions could be obtained by understanding
the cancellation mechanisms of scalar integrals. In the case of
a nontrivial tensor structure two--photon scattering in 
the gauged Yukawa model has been investigated. 
A result has been obtained in terms of
tensor coefficients by exploiting the Lorentz, gauge 
and analyticity structure as far as possible.
For the $++$ helicity amplitude a very compact representation
could be obtained. The $+-$ amplitude is 
presently under investigation. 

These results are steps towards a NLO description of $2\to 4$
processes at future colliders. For a realistic QCD amplitude
a lot of work remains to be done.
\section*{Acknowledgements}
I would like to thank the organizers of the RADCOR/Loops and Legs~2002
conference for their work and the pleasant and stimulating
atmosphere in Kloster Banz. 
I also want to thank 
J.~Ph.~Guillet, G.~Heinrich and C.~Schubert for giving 
me the opportunity to present some results of our collaboration.

\end{document}